\pdfoutput=1

\documentclass{nature_w_fig}

\usepackage{theorem,amsmath,enumerate,amssymb,amsfonts}
\usepackage{graphicx}
\usepackage[font=large]{caption}

\title{Graphene mechanical oscillators with tunable frequency}

\author{Changyao Chen$^{1, \dag}$, Sunwoo Lee$^{2,\dag}$, Vikram V. Deshpande$^3$, Gwan-Hyoung Lee$^{1,4}$, Michael Lekas$^2$, Kenneth Shepard$^2$, James Hone$^1$}

\begin{document}

\maketitle

\begin{affiliations}
 \item Department of Mechanical Engineering, Columbia University in the city of New York, USA
 \item Department of Electrical Engineering, Columbia University in the city of New York, USA
 \item Department of Physics, Columbia University in the city of New York, USA
 \item Samsung-SKKU Graphene Center (SSGC), Suwon, Gyeonggi 440-746, Korea
 \item[] \dag These authors contributed equally to this work
\end{affiliations}

\begin{abstract}
Oscillators, which produce continuous periodic signals from direct current power, are central to modern communications systems, with versatile applications such as timing references and frequency modulators\cite{Nguyen_ieee_2007,ekinci_rsi_2005,harjimiri_book_1999,Huang_nature_2003,Feng_nnano_2008,Villanueva_nl_2011,Hanay_nnano_2012}. However, conventional oscillators typically consist of macroscopic mechanical resonators such as quartz crystals, which require excessive off-chip space.  Here we report oscillators built on micron-size, atomically-thin graphene nanomechanical resonators, whose frequencies can be electrostatically tuned by as much as 14\%. The self-sustaining mechanical motion of the oscillators is generated and transduced at room temperature by simple electrical circuitry.  The prototype graphene voltage controlled oscillators exhibit frequency stability and modulation bandwidth sufficient for modulation of radio-frequency carrier signals. As a demonstration, we employ a graphene oscillator as the active element for frequency modulated signal generation, and achieve efficient audio signal transmission.\end{abstract}

Microscale microelectromechanical systems (MEMS) oscillators, which can be integrated on-chip, have demonstrated excellent frequency stability among other attributes\cite{Beek_jmm_2012}.  However, MEMS oscillators typically occupy large footprints on integrated circuits level, and achieve high frequency through large mechanical stiffness, which makes frequency tuning difficult. Therefore they are not well suited for use as voltage controlled oscillators (VCOs).  Nanoelectromechanical systems (NEMS)\cite{ekinci_rsi_2005} oscillators, on the other hand, can achieve high resonant frequencies\cite{Huang_nature_2003} while maintaining mechanical compliance needed for tunability, and only require small on-chip area.  Indeed, recent work has demonstrated NEMS oscillators at $>$ 400 MHz in SiC beams\cite{Feng_nnano_2008}, and at $\sim$14 MHz in AlN-based resonators\cite{Villanueva_nl_2011}. Both of these systems were designed for high frequency stability and low phase noise, as opposed to frequency tunability.  An additional challenge in NEMS is that their small size typically results in a small motional signal that can easily be overwhelmed by spurious coupling or background noise.  In the SiC oscillator, this problem was overcome using a cryogenic magnetomotive bridge technique, while in the AlN oscillators, parametric drive at twice the resonant frequency was used.

Graphene\cite{Geim_nmater_2007} is a material well suited for designing a NEMS VCO.  In particular, as an atomically thin ultrastiff and ultrastrong material\cite{Lee_Science_2008,Lee_science_2013}, it can achieve high resonant frequencies that can be externally tuned over a wide range with moderate applied voltage\cite{chen_nnano_2009,Zande_nl_2010,Song_nl_2011,eichler_nnano_2011,Singh_nano_2010}.  In addition, its charge-tunable conductance and large electrical mobility allows efficient transduction of mechanical vibration when a graphene membrane is configured as a suspended vibrating field effect device\cite{Nathanson_ieee_1967,xu_apl_2010}.  This allows direct radio-frequency (RF) electrical readout with signal to background ratios (SBR) larger than 10 dB at room temperature\cite{Lee_apl_2013}. In this letter, we report self-sustained graphene NEMS oscillators comprising a suspended graphene resonator and a simple electrical positive feedback loop. 

The graphene oscillator consists of suspended strip of chemical vapor deposited (CVD) graphene\cite{Li_Science_2009}, metal electrodes, and a clamping structure made from SU-8 epoxy photoresist that defines a circular graphene drum 2-4 $\mu$m in diameter. The graphene is  suspended over a metal local gate electrode on an insulating substrate, as shown in Fig 1a.  The SU-8 polymer clamping increases the mechanical rigidity of the suspended structure, allowing for a gate-channel spacing as small as 50 nm, and eliminates complex vibrational modes due to unclamped edges\cite{Lee_apl_2013}, without significantly degrading the electronic performance of underlying graphene: we observe field effect mobilities of up to 6,000 cm$^2$/Vs, similar to the devices without SU-8 support. The detailed fabrication process is described in Methods. Following fabrication, the mechanical resonance of each device is measured using a previously described technique\cite{Nathanson_ieee_1967,xu_apl_2010,Lee_apl_2013}.  Briefly, motion of the graphene is driven by applying a direct current (DC) and RF bias (combined with a bias tee) to the gate, and read out by applying a second DC bias to the drain. On resonance, the motion of the graphene modulates the charge density, which in turn modulates the conductance and drain current.  The electrical actuation and detection are performed using a vector network analyzer (VNA), which allows measurement of both the signal amplitude and phase. The large electronic mobility of graphene, combined with high mechanical compliance (spring constant ranges from 0.1 to 1 N/m) leads to efficient electro-mechanical coupling; small gate spacing (200 nm, equivalent to static capacitance of 44 aF/$\mu$m$^2$) and large sample size also contribute to high SBR even at room temperature\cite{Lee_apl_2013}, and facilitates direct electrical transduction of the mechanical motion.

\begin{figure}
\centering
\includegraphics[width=1.0\textwidth]{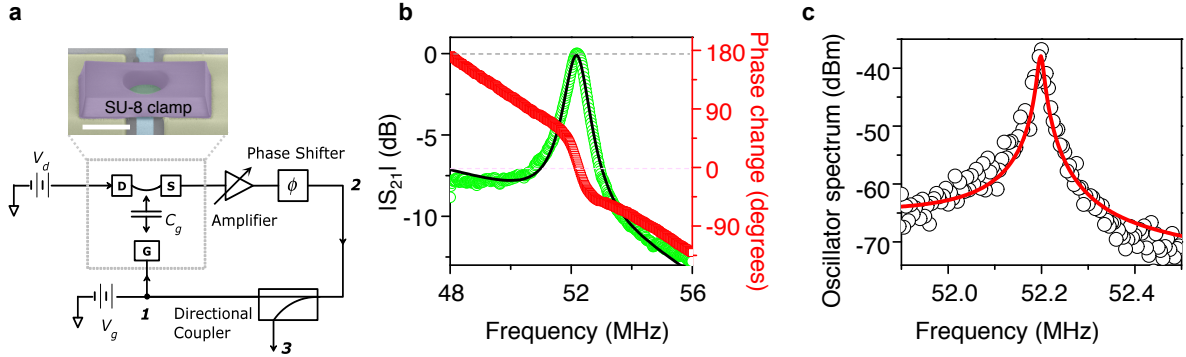}

\caption{\textbf{Self-sustained graphene mechanical oscillators.} \textbf{a}, Simplified circuit diagram of self-sustained graphene mechanical oscillators. See Supplementary Information section 1 for details. Inset: false-color SEM image of suspended graphene with circular SU-8 polymer clamp. Scale bar: 3 $\mu$m. \textbf{b}, Open-loop (resonator) transmission $S_\text{21}$ measurement for sample 1 (4 $\mu$m diameter drum, 200 nm vacuum gap), of both magnitude (green squares) and phase (red squares). A large resonant peak with 8 dB above a gradual decreasing background is observed. A corresponding Lorentzian fit is shown in black solid line, revealing a resonant frequency of 52.19 MHz and quality factor of 55. The magnitude and phase of $S_\text{21}$ are already set to satisfy Barkhausen criterion for self oscillation. Amplifier gain is 60 dB. \textbf{c}, Output power spectrum (black circles) of graphene mechanical oscillator for sample 1. The red solid line shows the forced Lorentzian fit as described in the main text. Both measurements at taken at room temperature, with $V_d$ = -0.5V, and $V_g$ = 10V. The driving power for resonator is -40 dBm.}
\end{figure}

To achieve self-oscillation, the system needs to satisfy the Barkhausen criterion\cite{harjimiri_book_1999}: the open-loop gain must be unity, and the feedback phase must be an integer multiple of 2$\pi$.  We perform open-loop characterization of the resonator by measuring the forward transmission, $S_\mathrm{21}$ between nodes 1 and 2 as shown in Fig. 1a. We then set the gain at resonance to unity with a variable gain amplifier, and adjust the phase to zero with a tunable phase shifter.  Figure 1b shows both the magnitude and phase of the measured $S_\mathrm{21}$ of sample 1 (4 $\mu$m diameter drum, 200 nm vacuum gap), with both of the Barkhausen criteria met. As the feedback loop is closed to generate self-oscillation, a 20 dB directional coupler is placed in the loop to monitor the oscillators's power spectrum and waveform in the time domain (node 3 in Fig 1a). Fig 1c shows the power spectrum of sample 1 under the same configuration shown in Fig 1b.  It shows clear self-oscillation at the open-loop resonant frequency.  

A distinctive signature of oscillators is the spectral linewidth compression compared to the corresponding passive resonators\cite{Ham_ieee_2003,Feng_nnano_2008}. The mechanisms of linewidth broadening in resonators and oscillators are inherently different: in resonators, the finite linewidth is due to the energy dissipation during each vibration cycle, and quantified by the quality factor, $Q$; in oscillators, the finite spectral linewidth is mostly due the phase noise\cite{Leeson_ieee_1966,Hajimiri_ieee_1998}, and quantified by the spectrum power density away from carrier frequency. Nevertheless, for the sake of direct comparison and without losing the generality, we use the full width at half maximum  (FWHM), $\Delta$, as the characteristic linewidth for both resonators and oscillators\cite{Feng_nnano_2008}. For sample 1 shown above, the resonator linewidth $\Delta^\mathrm{res} = f_0^\mathrm{res}/Q^\mathrm{res} \approx 935$ kHz, where $ f_0^\mathrm{res} = 52.19$ MHz is the resonant frequency and $Q^\mathrm{res} \approx$ 55. The oscillator has spectral linewidth of $\Delta^\mathrm{osc} =  f_0^\mathrm{osc}/Q_\mathrm{eff}^\mathrm{osc} \approx 13$ kHz, with oscillation frequency $ f_0^\mathrm{osc}$ = 52.20 MHz, and an effective quality factor $Q_\mathrm{eff}^\mathrm{osc} \approx$ 4,015.  The linewidth compression ratio $\Delta^\mathrm{osc}/\Delta^\mathrm{res}$ is 72 in this case. We have observed that the oscillator power spectrum is sensitive to feedback loop gain and phase, which can modify apparent $ f_0^\mathrm{osc}$ and $\Delta^\mathrm{osc}$ (see Supplemental Information section 2 and 3).

\begin{figure}
\centering
\includegraphics{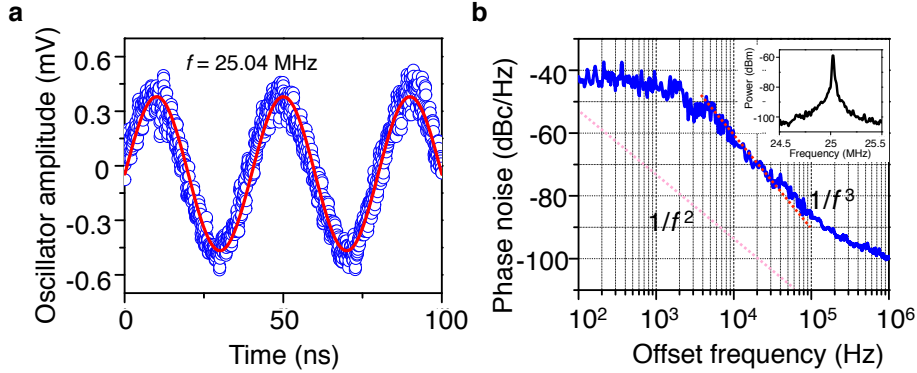}

\caption{\textbf{Stability of graphene mechanical oscillators.} \textbf{a}, Time-domain waveform (blue circles) of graphene mechanical oscillation acquired by digital oscilloscope. The red solid line is sinusoidal fit with oscillation frequency of 25.04 MHz. \textbf{b}, Phase noise as function of offset frequency from carrier. The plateau in phase noise up to 2 kHz is due to Lorentzian-nature broadening of the embedded resonator. For offset frequency above 2 kHz, the phase noise shows roughly a 1/$f^3$ dependence (red solid line). The pink dashed line is the phase noise limit due to the thermal fluctuation. The inset shows the corresponding power spectrum. Both data are acquired from sample 2 (2 $\mu$m diameter drum, 200 nm vacuum gap) at room temperature with $V_d$ = 0.3V and $V_g$ = -5V. }
\end{figure}

In Fig. 2a, we show the time domain response of sample 2 (2 $\mu$m diameter drum, 200 nm vacuum gap), displaying clear sinusoidal waveform with peak-to-peak amplitude of 0.8 mV, corresponds to vibrational amplitude of 0.14 nm. The corresponding phase noise is shown in Fig. 2b: it is constant up to 2 kHz offset frequency, then decreases with a $1/f^3$ slope.  The origin of the flat plateau is the Lorentzian-nature linewidth broadening of the resonator\cite{Ham_ieee_2003} from white noise coupling into the gate, and the $1/f^3$ dependence is due to $1/f$ (flicker) noise. Interestingly, we do not observe the expected $1/f^2$ contribution from thermal (white) noise, which indicates that the stability of our graphene mechanical oscillation is still limited by external circuitry (for example, the DC gate voltage source and feedback amplifiers). To estimate the potential for improvement, we calculate the intrinsic phase noise due to thermal sources given by\cite{Leeson_ieee_1966,Yang_thesis_2006}: $\mathfrak L (f) = 10 \log [({k_\text{B}T{f_0^\text{res}}^2})/({2P_C {Q^\text{res}}^2f^2})]$, where $f$ is the offset frequency, $k_\text{B}$ is the Boltzmann constant, $T$ is the temperature and $P_C$ is the carrier power. The intrinsic phase noise of sample 2 ($P_C \approx$ 126 nW, and $Q^\text{res} \approx$ 15) is shown as the dashed line in Fig 2b.   The intrinsic phase noise at an offset frequency of 1 kHz equals to -73 dBc/Hz, which is more than two orders of magnitude smaller than observed value. Graphene oscillator built from samples with higher open-loop $Q$ do not show improved phase noise (Supplementary Information Fig. S11), further indicating that different processes are responsible for linewidth broadening in the open-loop and closed-loop configuration. Furthermore, since the closed-loop oscillators are running inside the nonlinear regime, it is possible to evade the amplifier noises by setting the open-loop condition at special operation points\cite{Villanueva_prl_2013}. 
 
\begin{figure}
\centering
\includegraphics{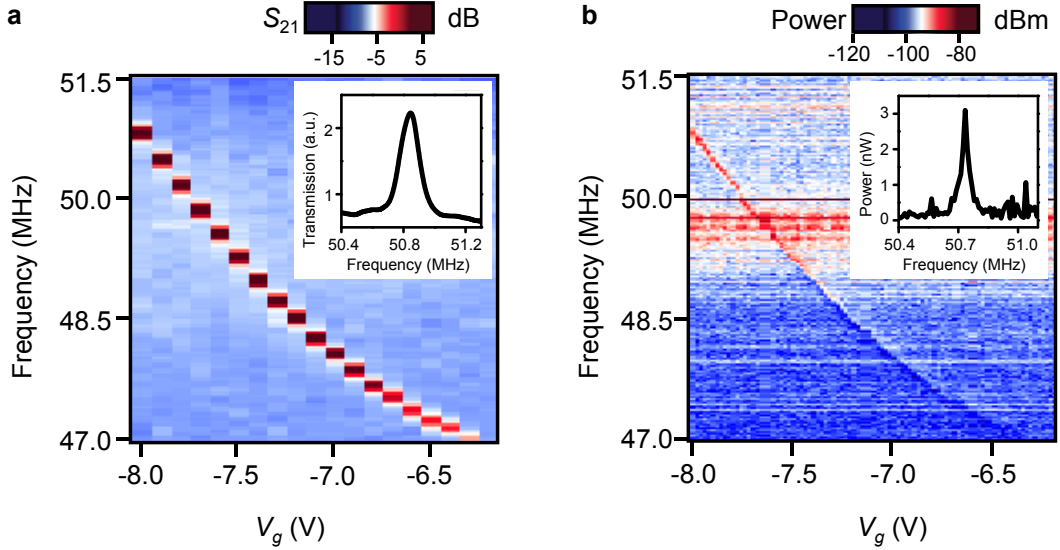}

\caption{\textbf{Voltage controlled tunable oscillations.} \textbf{a}, Open-loop transmission $S_\text{21}$ measurement. $S_\text{21}$ is shown in color plot as function of both actuation frequency and applied gate voltage $V_g$. The tuning sensitivity is about 2.7 MHz/V over the $V_g$ range shown. Inset: Open-loop transmission in linear scale for $V_g$ = -8V. \textbf{b}, Corresponding oscillation power spectrum under the same condition, showing close resemblance of frequency tuning. Both data are acquired from sample 3 at 77 K, with $V_d$ = - 0.1V and the driving power of resonator is -55 dBm. Inset: Power spectrum in linear scale for $V_g$ = -8V. }
\end{figure}

Because graphene is atomically thin, its resonant frequency is dominated by in-plane tension, which can be modified electrostatically by applying a DC voltage $V_g$ to the back gate.  The degree of tunability depends on the initial built-in tension\cite{Singh_nano_2010,Song_nl_2011,Zande_nl_2010}, and can reach 400\% with lowest built-in tension\cite{chen_nnano_2009}  (the devices used here typically show much  smaller tunability due to tension imparted by the SU-8 clamps\cite{Lee_apl_2013}). The same tuning mechanism can be readily used to realize highly tunable VCOs. Figure 3a shows open-loop characterization of sample 3 (no SU-8 support, 4.2 $\mu$m long and 2 $\mu$m wide, 200 nm vacuum gap):  in applied $V_g$ range from -8 V to -6.2 V, we observe clear resonance tuning from 51.5 MHz to 47 MHz. After we close the positive feedback loop, we find similar oscillation frequency tuning with $V_g$ faithfully follows the dependence, as shown in Fig 3b.  Notably, the tuning sensitivity for this sample is about 2.7 MHz/V, comparable to commercial very high frequency (VHF) VCOs; devices with lower built-in tension showed tuning sensitivity up to 8 MHz/V (Supplementary Information section 4).  We note that the tuning range in this work is limited by the external feedback circuit, which introduces extra phase shifts during the frequency tuning to violate the Barkhausen criteria.  This can be overcome by manual adjustment of phase delay at individual operation point: we demonstrate tuning up to 14\% in this way (Supplementary Information Fig. S12). On-chip readout circuitry or software-controlled phase compensation should largely eliminate this tuning range limitation. For comparison, the tunability of commercial very high frequency (VHF) VCOs  available from Minicircuits ranges from 0.11\% to 106\%, with corresponding tunability from 1.5 MHz/V to 161 MHz/V. 

NEMS hold promise for mechanical RF signal processing, as elements such as filters, modulators, and mixers. In fact, previous work has demonstrated radio receivers based on both carbon nanotubes and Si-NEMS\cite{Jensen_nl_2007,Bartsch_nano_2012}.  Here we demonstrate the use of a graphene VCO to create the complementary structure -- a NEMS radio transmitter, which up-converts an audio signal into a frequency-modulated (FM) carrier signal. Graphene VCOs are well suited for this application: their oscillation frequencies can be tuned into the FM broadcast band (87.7 MHz to 108 MHz) with proper device geometry design. Moreover, the modulation bandwidth, which quantifies how rapidly the VCO can respond to changes in tuning voltage, is sufficient for audio signals (above 15 kHz in our prototype, currently limited by the cut-off frequency of DC port in the bias tee used in the test circuitry).  In the demonstration, as depicted in Fig. 4a (sample 4, 3 $\mu$m diameter drum, 200 nm vacuum gap), the audio signal is added to the DC gate bias, modulating the output frequency and generating an FM signal at the carrier frequency ($\sim$100 MHz).  We then feed the signal directly into a commercial FM receiver to recover the original audio signal.  Before we send in complex multi-frequency signals, we first transmit a pure sinusoid signal at 440 Hz, and acquire the down-coverted signal with digital oscilloscope, as shown in the insert of Fig. 4a. The total harmonic distortion is 9.4\% up to the 20$^\text{th}$ harmonic. Next, we transmit a more complex music waveform, and connect the de-modulated output directly to a speaker. Fig 4b and Fig 4c show 1 second segments of original and received audio waveforms, of ``Gangnam Style'' by PSY. It is clear that we recover a signal that faithfully reproduces the original. The longer version of this song can be found in Supplementary Information. During the test, when we purposely detune the frequency of the FM receiver, the sound is lost, confirming the validity of our graphene VCO operation.  

\begin{figure}
\centering
\includegraphics[width=0.6\textwidth]{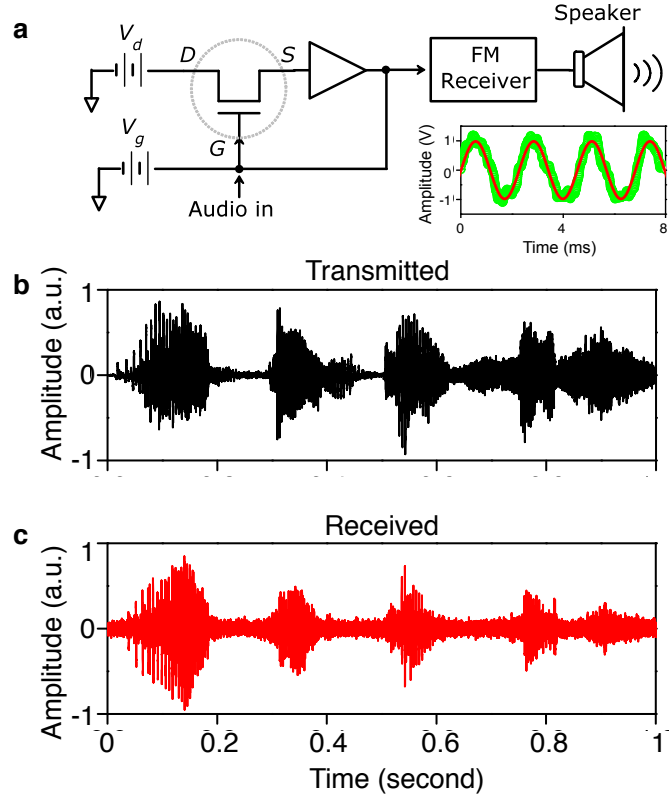}

\caption{\textbf{Graphene radio station.} \textbf{a}, Simplified circuit diagram of graphene radio station. Inset shows the received 440 Hz signal (green circles) after FM demodulation. The red line is sinusoidal fit. \textbf{b}, Audio waveform of 1 second soundtrack of ``Gangnam Style"  by PSY, transmitted through the graphene radio station. \textbf{c}, Audio waveform of the received soundtrack by FM receiver, faithfully reproduced the original audio signal and can be heard clearly (see Supplementary Information). Sample (3 $\mu$m diameter drum, 200 nm vacuum gap) is tested at room temperature, with $V_d$ = -1.2V and $V_g$ = -4.5V, oscillation frequency is 100 MHz.}
\end{figure}

In conclusion, we have demonstrated self-sustained graphene mechanical oscillators with tunable frequency operation at room temperature. The high quality CVD graphene used in this study indicates the possibility of wafer-scale integration of graphene resonant NEMS, which is compatible with current CMOS fabrication processes\cite{bae_nnano_2010}. Beyond the graphene �radio station� shown above, there are many immediate applications that can utilize nano-scale, tunable VCOs\cite{Nguyen_ICSSAM_2005}, such as \textit{in situ} mass sensing and RF signal processing\cite{Nguyen_pieee_1998}, and noise suppression with frequency synchronization\cite{Cross_prl_2004}. The work described above has clearly demonstrated the promise of graphene-based NEMS, and opens new regimes for miniatured size NEMS-based large-scale electronic circuitry.


\begin{methods}
\subsection{Sample fabrication}
All samples, except for sample 3, are derived from CVD graphene grown on copper foil substrates\cite{Li_Science_2009}. We transfer the CVD graphene to pre-patterned substrates made from high-resistivity silicon, with gate electrodes buried under plasma-enhanced chemical vapor deposition (PECVD) oxide. The PECVD oxide was planarized with chemical mechanical polishing (CMP), in order to promote the adhesion between the CVD graphene and the substrate.  After patterning source (S), drain (D) electrodes and SU-8 polymer for circular clamping with electron beam lithography, we immerse the whole sample into buffered oxide etchant (BOE) to release the suspended graphene drum resonators\cite{Lee_apl_2013}.  Vacuum gap between graphene and underneath local gate is controlled by PECVD oxide thickness and CMP duration: we can achieve the vacuum gap from 50 nm to 200 nm, and fabrication yield greater than 70\% for suspending graphene. To fabricate sample 3, we directly exfoliated graphene onto pre-patterned electrodes / trenches structure as we reported previously\cite{xu_apl_2010}.

\subsection{Oscillator characterization}
All experiments are carried out in a high-vacuum ($<$ 10$^{-5}$ Torr) probe station. We only choose samples with large open-loop SBR ($>$ 5 dB) to construct graphene mechanical oscillators.  To adjust the feedback phase and gain, we use phase shifters (Lorch Microwave) and a tunable amplifier (Mini-circuits ZFL-1000G). Upon confirming that the open-loop gain is unity and total phase shift is multiple of 2$\pi$, we close the loop by connecting node 1 and 2 (as shown in Fig. 1a). The completed circuit diagram can be found in Supplemental Information section 1.

Closed-loop measurements are performed with spectrum analyzer (Agilent E4440A) for both spectral characterization and phase noise measurement (option 226). The time domain data are acquired by Agilent mixed signal oscilloscope (MSO-X 2014A).

In modulation bandwidth test, we add a square-wave with 0.4 V peak-to-peak value for modulation (Stanford Research System DS345).  The applied modulation frequency is from 1 Hz to 100 kHz. The DC voltage and low frequency modulation signal are combined with summing amplifier (Stanford Research System SIM 980), and then applied to the DC port of the bias tee while the RF excitation is applied to the RF port.

The measurement setup of FM transmission is very similar to that of the modulation bandwidth test. Instead of the square-wave, we apply audio signal to the summing amplifier, and graphene acts as both oscillator and mixer, allowing for FM transmission.  The modulated signal is then transmitted to the standard radio receiver (NAD Stereo Tuner 4220) where the sounds signal is demodulated before played through a speaker.

\end{methods}

\begin{addendum}
 \item The authors like to thank Philip Kim, John Kymissis, Arend van der Zande, Nicholas Petrone, Alexander Gondarenko, Eugene Hwang, Changhyuk Lee, Alyosha Molnar, and Victor Abramsky for critical discussions.  Fabrication was performed at the Cornell Nano-Scale Facility, a member of the National Nanotechnology Infrastructure Network, which is supported by the National Science Foundation (Grant ECS-0335765), and Center for Engineering and Physical Science Research (CEPSR) Clean Room at Columba University.  The authors acknowledge support by Qualcomm Innovation Fellowship (QInF) 2012 and AFOSR MURI FA9550-09-1-0705. \item[Competing Interests] The authors declare that they have no competing financial interests.

\item[Author contributions] C.C. and S.L. and J.H. conceived and designed the experiments, C.C., S.L. and V.V.D. performed the experiments and analyzed the data, G-H.L. provided the samples, M.L. and K.S. contributed measurement/analysis tools, C.C., S.L. and J.H. co-wrote the paper. All authors discussed the results and commented on the manuscript.
 \item[Correspondence] Correspondence and requests for materials
should be addressed to \\
J.H.~(email: jh2228@columbia.edu).

\item[Additional information] Supplementary information accompanies this paper at\\ www.nature.com/naturenanotechnology. Reprints and permission information is available online at\\ http://npg.nature.com/reprintsandpermissions/.
Correspondence and requests for materials should be addressed to J.H..
\end{addendum}



\end{document}